\documentclass[preprint, 5p,times, twocolumn]{elsarticle}

\usepackage{graphicx}
\usepackage{epsfig}

\usepackage{bm}

\usepackage{slashed}
\usepackage[usenames,dvipsnames]{color}
\usepackage[normalem]{ulem}
\usepackage{comment}
\usepackage{latexsym}
\usepackage{natbib}
\usepackage{url}
\usepackage{dcolumn}
\usepackage{color}
\usepackage{amsfonts,amssymb,amsmath}
\usepackage{graphicx,epsfig}
\usepackage{psfrag}
\usepackage{subfigure}
\usepackage{hyperref}
\hypersetup{colorlinks=true}

\journal{Physics Letters B}

\begin{document}

\begin{frontmatter}
\title{\boldmath  Higgs instability and de-Sitter radiation}

\author [prl,iet]{Gaurav Goswami}
\ead {gaurav.goswami@ahduni.edu.in}

\author [prl]{Subhendra Mohanty}
\ead {mohanty@prl.res.in}

\address [prl]{ Physical Research Laboratory, Navrangpura, Ahmedabad 380 009, India.}
\date{\today}
\address [iet]{Institute of Engineering and Technology, Ahmedabad University, Navrangpura, Ahmedabad 380 009, India.}

\begin{abstract}
If the Standard Model (SM) of elementary particle physics is assumed to hold good to arbitrarily high energies, then,
for the best fit values of the parameters, 
the scalar potential of the Standard Model Higgs field turns negative at a high scale $\mu_{\rm inst}$. 
If the physics beyond the SM 
is such that it does not modify this feature of the Higgs potential and if the Hubble parameter during inflation 
($H_{\rm inf}$) is such that 
$H_{\rm inf} \gg \mu_{\rm inst}$, then,  quantum fluctuations of the SM Higgs during 
inflation make it extremely unlikely that after inflation it will be found in the metastable vacuum at the weak scale.
In this work, we assume that 
(i) during inflation, the SM Higgs is in Bunch-Davies vacuum state, and, 
(ii) the question about the stability of the effective potential must be answered in the frame of the freely falling observer 
(just like in Minkowski spacetime), and then use the well known fact that the freely falling observer finds Bunch-Davies vacuum to be in thermal state to show that the probability to end up in the electroweak vacuum 
after inflation is reasonably high.
\end{abstract}


\end{frontmatter}

\begin{figure}
  \begin{center}
    \includegraphics[width=.45\textwidth]{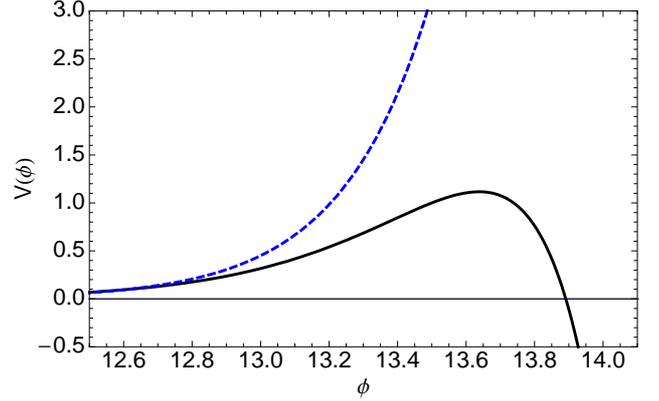}
  \end{center}
\caption{In this cartoon fig., the solid curve is the shape of quantum effective potential of SM Higgs field
around the field values of the order of $\phi \sim \mu_{\rm inst}$ (the numbers in the fig are in arbitrary units). 
In the present work, we show that during inflation,  
the Gibbons-Hawking temperature of quasi-de-Sitter space ensures that the field 
is in thermal state with the corresponding thermal effective potential which qualitatively looks like the 
dashed curve above. This means the thermal effects shall the Higgs instability problem without requiring 
any other new physics below $\mu_{\rm inst}$ pushing the effective instability scale to a much higher value.}
\label{illustration}
\end{figure}

Various observations and theoretical considerations indicate that there exists 
physics beyond the Standard Model (SM), but it is unclear at what scale this new physics exists. 
Renormalization group evolution of SM couplings shows that the Higgs quartic coupling 
becomes negative at a scale $\mu_{\rm inst} \approx 10^{10}-10^{12}$ GeV if 
there is no new physics beyond the SM
\cite{2012JHEP...02..037H,2012PhLB..709..222E,2012PhRvD..86a3013X,2012JHEP...06..033C,2012JHEP...10..140B,
2012JHEP...08..098D},
\cite{2013JHEP...12..089B},
\cite{Spencer-Smith:2014woa,DiLuzio:2014bua,Andreassen:2014eha,Andreassen:2014gha}.
This implies that at large field values, the quantum effective potential of SM Higgs field must look like 
the solid curve in Fig (\ref{illustration}) rendering the electroweak vacuum metastable 
(i.e. the corresponding lifetime turns out to be bigger than the age of the universe, 
however, see \cite{Branchina:2013jra}).

The energy scale of inflation is unknown at present. Interpreting the recent BICEP2 
observations of B-mode polarization of cosmic microwave background radiation 
\cite{Ade:2014xna} as being due to inflationary gravitational waves,
one infers the Hubble parameter during inflation to be $H_{\rm inf} \sim 10^{14}$ GeV. 
However, the signal observed by BICEP2 experiment is best interpreted as being due to cosmic dust 
\cite{2014arXiv1409.5738P,2014arXiv1409.4491C} so that the energy scale of inflation
continues to be unknown at present and can be potentially found by future CMB experiments. 
If however the Hubble parameter during inflation is larger than the scale 
$\mu_{\rm inst} \approx 10^{10}-10^{12}$ GeV,
the quantum fluctuations of the SM Higgs during inflation shall cause it 
to run away to the global minimum in the effective potential at very high field values 
leaving no way of reaching the local minimum at $\phi \approx 250$ GeV, the electroweak vacuum.
This is why it is often argued that
\cite{2008JCAP...05..002E,2013PhLB..719..415L,2013PhLB..722..130K,2014PhRvL.112t1801F,
2014JCAP...07..025E,2014arXiv1404.4709K,2014arXiv1404.5953H,2014PhRvL.113u1102H,
Kamada:2014ufa, Shkerin:2015exa,Kearney:2015vba,Espinosa:2015qea}
considerations of inflationary cosmology 
imply that if $\mu_{\rm inst} < H_{\rm inf}$ (the Hubble parameter during inflation), 
new physics must show up at some energy scale below $\mu_{\rm inst}$ (see also \cite{Kusenko:2014lra}).

Assuming that the SM Higgs is not the inflaton 
(unlike \cite{Masina:2011un} and \cite{Bezrukov:2014ipa}), the inflaton must belong to a beyond-SM (BSM) scenario. 
Often, the SM Higgs potential in the BSM scenario gets modified such that there is no Higgs instability 
problem. E.g. in \cite{Lebedev:2012zw}, it is argued that if the quartic coupling of inflaton and Higgs 
is greater than $10^{-11}$, then too the problem is avoided. It is thus very important for the very existence 
of this problem that the BSM scenario is such that it modifies the Higgs potential only very slightly.

For a quantum field in inflationary quasi-de-Sitter spacetime in Bunch Davies vacuum state,
every freely falling (i.e. geodesic) observer is surrounded by thermal radiation with the 
Gibbons-Hawking temperature of $\frac{H}{2\pi}$ (where $H$ is the Hubble 
parameter during inflation) 
\cite{PhysRevD.15.2738,PhysRevD.37.354,RevModPhys.57.1,2001hep.th...10007S,2006JHEP...04..057G}. 
As we argue below, the SM Higgs field is expected to be in Bunch-Davies vacuum and then the geodesic 
observers must find it in a thermal state.
In such a scenario, to answer any questions related to the dynamics of the SM Higgs field 
and hence of the stability of the electroweak vacuum, one should analyse the corresponding 
thermal effective potential of the Higgs field. 
We found that the thermal effective potential of the SM Higgs field during inflation 
is such that in the post inflationary universe, the fraction of Hubble volumes found with Higgs displaced 
such that it can reach the electroweak vacuum is quite significant.
Thus, no new physics is needed below $\mu_{\rm inst} \approx 10^{10}-10^{12}$ GeV
to ensure that the universe after inflation ends up in the local minimum at 
$\phi \sim 250$ GeV, the electroweak vacuum. 
But, in order to have inflation at $\mu \sim V_{\rm inf}^{1/4}$, one still expects some new 
physics to turn up at an energy scale below $V_{\rm inf}^{1/4} \approx 10^{16}$ GeV.
As we shall see, we actually need new physics at a scale below $H_{\inf} \approx 10^{14}$ GeV.

It is worth emphasising that all the proposed solutions of this problem (see e.g. \cite{2012JHEP...06..031E})
assume either the existence of physics beyond the standard model at an energy scale below
$\mu_{\rm inst}$, or assume that gravity is described by a theory other than Einstein's GR
(often a scalar-tensor theory \cite{2014PhRvL.113u1102H} in which the SM Higgs field acts as the 
scalar degree of freedom).
In this work on the other hand, as we argue below, the considerations of well established physics 
(Einstein's GR and basic QFT in curved spacetime) in fact imply that the problem becomes insignificant, 
provided one takes into account the phenomenon of Hawking-Gibbons temperature. 
Recall that the existence of de-Sitter radiation (the equivalent of Hawking radiation in de-Sitter space) 
is an inevitable consequence of QFT in curved spacetime and has been well known and established for 
decades.

We begin 
by recalling the argument in favour of the hypothesis that 
$\mu_{\rm inst} < H_{\rm inf}$ implies new physics below $\mu_{\rm inst}$. 
Then, after reminding why there must be thermal radiation 
in inflationary de-Sitter spacetime, we evaluate the quantum effective potential of SM 
Higgs and then show how the corresponding thermal effective potential of the SM Higgs 
helps. 
We then conclude 
with a summary and possible issues.

{\it Cosmic inflation and Higgs instability}:
In the Standard Model of elementary particle physics, the one-loop beta function of the self coupling 
$\lambda$ of Higgs receives a negative contribution from the loop of the top quark while 
it receives positive contribution from the Higgs loop. 
A heavy top quark and a light Higgs thus ensure that as we probe higher energies, 
at some scale $\mu_{\rm inst} \approx 10^{10}-10^{12}$ GeV, $\lambda$ becomes zero 
and eventually negative at even higher energies
\cite{2012JHEP...02..037H,2012PhLB..709..222E,2012PhRvD..86a3013X,2012JHEP...06..033C,2012JHEP...10..140B,
2012JHEP...08..098D},
\cite{2013JHEP...12..089B},
\cite{Spencer-Smith:2014woa,DiLuzio:2014bua}.
The uncertainties in the value of this scale are determined predominantly by the uncertainties in the 
measured value of the mass of top quark. 

If the recent observations of BICEP2 \cite{Ade:2014xna} collaboration are to be interpreted as 
being due to the inflationary gravitational waves, it appears that the inferred energy scale of inflation 
\cite{PhysRevLett.73.3347} is 
\begin{equation}
V^{1/4}_{\rm inf} \approx 2.2 \times 10^{16} {\rm GeV} \left( \frac{r}{0.2} \right)^{1/4} \; ,
\end{equation}
with $r = {\cal O}(0.1$) and hence
\begin{equation} \label{eq:H-r}
 H_{\rm inf} \approx 1.2 \times 10^{14} {\rm GeV} \left( \frac{r}{0.2} \right)^{1/2} \; . 
\end{equation}
Even if the best explanation of BICEP2 observations is in terms of dust
\cite{2014arXiv1409.5738P,2014arXiv1409.4491C}, 
an energy scale of inflation of this order has triggered arguments 
\cite{2008JCAP...05..002E,2013PhLB..719..415L,2013PhLB..722..130K,
2014PhRvL.112t1801F,2014JCAP...07..025E,2014arXiv1404.4709K,2014arXiv1404.5953H}
purely from inflationary cosmological considerations, that
there must be new physics below the scale $\mu_{\rm inst}$. 
These arguments are based on the following reasoning:
since $\lambda$ turns negative at $\mu_{\rm inst}$, the quantum effective potential of the Standard Model 
Higgs field must look like the solid curve in Fig (\ref{illustration}). 
For any massless (i.e. sufficiently light) canonically normalized scalar field on quasi-de-Sitter background, 
every Fourier mode has, at late times, a quantum fluctuation of approximately $H^2$
(see e.g. \cite{mukhanov2005physical,Baumann_TASI-2009} for details) i.e.
\begin{equation}
\lim_{t \rightarrow \infty} \langle 0 | \tilde{\phi}(t,{\bf k}) \cdot \tilde{\phi}(t,{\bf k}') | 0 \rangle = 
 \delta^3({\bf k} + {\bf k}') \frac{2 \pi^2}{k^3} \left[ \frac{ H(t_k)}{2 \pi } \right]^2 \; .
\end{equation}
Here, $\tilde{\phi}$ is the three dimensionful Fourier transform of the field $\phi$ (so that the mass dimension of 
$\tilde{\phi}$ is -2), $H$ is the Hubble parameter during inflationary quasi-de-Sitter phase, 
$t_k$ is the time when the mode in question crosses the then Hubble radius and the state $| 0 \rangle$ is
the Bunch-Davies vacuum.

Thus, in inflationary quasi de-Sitter spacetime, at every scale, there is quantum fluctuation of the order of $H$. 
This shall happen for every light field during inflation including the Standard Model Higgs field itself.
Suppose (as the data suggests) $H_{\rm inf} > \mu_{\rm inst}$, this would then imply that just due to quantum 
fluctuations, averaged over a box of any size, the Standard Model Higgs field is going to be found 
in the extreme right portion of the effective potential (the solid curve) in Fig (\ref{illustration}).
Thus, during inflation, the large inflationary energy density can drive the Higgs out of electroweak vacuum 
i.e. the likelihood that Higgs field fluctuates to the unstable region of the potential is sizeable, even if 
Higgs begins inflation in EW vacuum.
The probability to have a Universe at the end of inflation which 
survived the quantum Higgs fluctuations is quite low \cite{2008JCAP...05..002E}.

This causes the field to runaway to even higher values and at the end of inflation we never end up in 
the desired SM electroweak vacuum at $\phi \sim 250$ GeV.
How did the universe end up in such an energetically disfavored state as the present electroweak vacuum?
Moreover, as the SM Higgs rolls down along the run away region of its effective potential, 
its negative energy density keeps on increasing until there comes a moment when 
it overpowers the energy density of inflaton itself, a process which can disrupt inflation. 
In \cite{2014JCAP...07..025E}, the authors argued that since in the SM, for the best fit value of the 
mass of top quark, the value of $V(\phi_{\rm max})$ (the potential energy at the local maximum) is 
lesser than $H^4_{\rm inf}$ (with $H_{\rm inf}$ assumed to be ${\cal O}(10^{14})$ GeV), inflationary 
fluctuations shall push it to $\phi > \phi_{\rm max}$ region of field space and hence new physics shall 
be required to modify the Higgs potential and make it stable against inflationary fluctuations.
In general, it is often argued that this means that there must be new physics at energy scale below 
$\mu_{\rm inst}$ which modifies the Higgs potential so that after inflation, we end up being in the 
correct vacuum (see e.g. \cite{2012JHEP...06..031E} for an example of new physics which could cure 
this problem).

Next, we show that the considerations of Gibbons-Hawking temperature
in quasi-de-Sitter background during inflation shall cause the corresponding thermal effective 
potential of the Higgs field to be of the form of the dashed curve shown in Fig (\ref{illustration})
suggesting that the above conclusion about the instability of the Standard Model Higgs during inflation
\cite{2008JCAP...05..002E,2013PhLB..719..415L,2013PhLB..722..130K,2014PhRvL.112t1801F,
2014JCAP...07..025E,2014arXiv1404.4709K,2014arXiv1404.5953H}
is not correct.


{\it Effects of de-Sitter radiation}:
Consider a free massless (or light) quantum scalar field in inflationary (quasi) de-Sitter spacetime.
It is known that a state which an observer using conformal (i.e. planar) coordinates describes as 
Bunch-Davies vacuum, to an observer using static coordinates, shall appear 
to have particles and these particles have a thermal spectrum, the corresponding temperature being 
$T = H/2\pi$ in natural units (see \cite{PhysRevD.15.2738,PhysRevD.37.354} for the original reference 
and sec V of \cite{RevModPhys.57.1} and \cite{2006JHEP...04..057G} for a review).
In fact, any geodesic observer moving along a timelike geodesic in 
de Sitter space observes a thermal bath of particles when the scalar field is in the Bunch-Davies vacuum
(see \cite{2001hep.th...10007S} for a review).

We posit that, during inflation, the SM Higgs must be in the Bunch Davies vacuum state because 
of the reason that at early times, the physical wavelength of any given mode is arbitrarily 
short compared to Hubble length and so, the distinction between Minkowski space and de-Sitter space 
must be negligible \cite{mukhanov2007introduction}.
It is well known that no de-Sitter invariant vacuum exists for an exactly massless scalar field
\cite{1985PhRvD..32.3136A}, but the SM Higgs has a small but non-zero mass. 
It is worth emphasizing that the choice of Bunch Davies vacuum state is also de-Sitter invariant
(see \cite{2006JHEP...04..057G} for a discussion).
In Minkowski spacetime, we analyse the questions of stability of effective potential in an inertial 
frame which is a freely falling frame. So, in this work, we take the point of view that in inflationary
quasi-de-Sitter spacetime, the observer in whose frame the questions of stability of the effective 
potential must be analysed is a freely falling observer, i.e. the one who uses static coordinates 
and who therefore detects de-Sitter radiation.
Since the temperature of de-Sitter radiation does not drop as the universe inflates, 
the dynamics of SM Higgs and the stability of the EW vacuum must be determined by its 
thermal effective potential. For this reason, we now find the thermal effective potential of SM Higgs 
at a temperature of $H_{\rm inf}$ and analyse its stability.


Let us find the one-loop quantum effective potential of the Standard Model Higgs.
The Higgs potential is $V = - \frac{m^2 |\varphi|^2}{2} + \lambda |\varphi|^4$ for 
the Higgs doublet $\varphi$ defined by 
\[ \left( \begin{array}{c}
\chi \\
 (\phi + i \eta)/\sqrt{2} \end{array} \right)\] 
where $\phi$ is the rolling physical Higgs field and $\eta$ and $\chi$ are the neutral and charged
Goldstones respectively.
Recall that $m^2$ is the only dimensionful parameter in the Standard Model Lagrangian.
We find the one-loop effective potential of the Higgs field due to its interactions with itself, 
with gauge bosons and with the top quark (all the other couplings are neglibly small).
The quantum effective potential can be rewritten as its tree level expression 
\begin{equation}
 V_{\rm eff} = - \frac{m (\mu) ^2 \phi^2}{2} + \lambda (\mu) \phi^4 \; ,
\end{equation}
but with the renormalized couplings (and with the renormalization scale $\mu$ set to $\phi$).
We thus need to find the Renormalization Group (RG) evolution of the various parameters
and couplings.
The RG flow of $m^2$ and $\lambda$ is determined approximately by 
the three gauge couplings $g_1$, $g_2$, $g_3$ and by the Yukawa coupling of top quark $y_t$.
One can easily solve the RGEs \cite{2013JHEP...12..089B} for the 6 parameters $(g_1, g_2, g_3, y_t, \lambda, m^2)$, 
with the values of these parameters at an initial renormalization scale (which we take to be $M_t = 173.1$ GeV)
chosen to be $(0.461,0.648,1.166,0.936,0.127,(132.7~{\rm GeV})^2)$ respectively \cite{2013JHEP...12..089B}
\footnote{Notice that all the parameters that we are working with are the ones defined in $\overline{MS}$ renormalization 
scheme and hence are gauge-invariant (see \cite{2013JHEP...12..089B} for the proof and relevant literature).
}.
We truncate our computation at one loop accuracy since our aim is to only illustrate that the thermal effects solve 
the problem we addressed in the last section.

\begin{figure}
  \begin{center}
    \includegraphics[width=.45\textwidth]{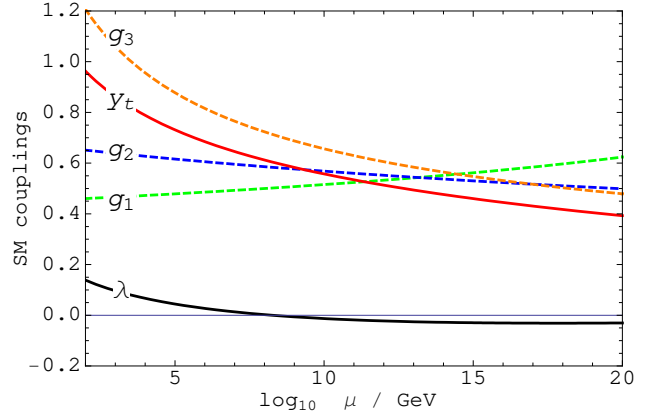}
  \end{center}
\caption{The RG evolution of SM couplings (defined in $\overline{MS}$ renormalization scheme) in one loop approximation. 
Note the approximate unification of gauge couplings around $10^{14}$ GeV and also
notice that $\lambda$ turns negative at an energy scale of $10^{8}$ GeV (since we have truncated the computation at 
one loop). }
\label{coupling_zero}
\end{figure}

\begin{figure}
  \begin{center}
    \includegraphics[width=.45\textwidth]{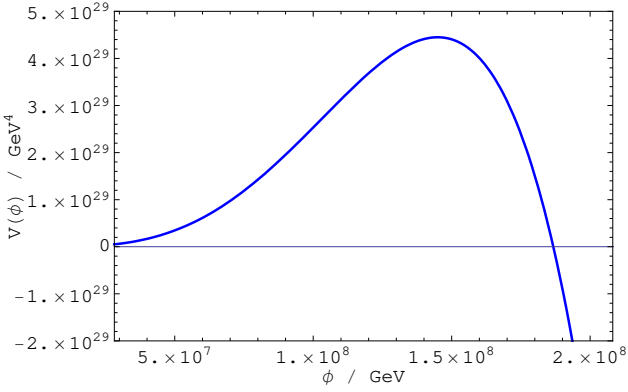}
  \end{center}
\caption{
In one loop approximation, the RG evolution of $\lambda$ causes it to become negative at 
an energy scale of around $10^{8}$ GeV. 
This causes the one-loop zero temperature quantum effective potential of 
the SM Higgs field to look like what is shown here. 
The quantum fluctuations of the SM Higgs field 
during inflation cause it to runaway to large field values spoiling any chances of getting the 
familiar low energy phenomenology after inflation.}
\label{potential_zero}
\end{figure}

We find that the couplings flow as shown in Fig (\ref{coupling_zero}), it can be seen that the 
self coupling of the SM Higgs becomes negative at a high energy scale. 
This scale is $10^8$ GeV in Fig (\ref{coupling_zero}) instead of 
the value $\mu_{\rm inst} \approx 10^{11}$ GeV (see \cite{2013JHEP...12..089B}) 
as we have truncated the RGEs at one loop approximation.
In Fig (\ref{potential_zero}) is plotted the corresponding quantum effective potential.
Around $10^8$ GeV, the corresponding Higgs potential drops below its value for the EW vacuum as 
shown in Fig (\ref{potential_zero}).
This illustrates the Higgs instabiliy problem at zero temperature.

In a thermal state, 
all the averages shall be ensemble averages of statistical fluctuations and not the averages 
over quantum fluctuations.
Thus, unlike the cases in which we wish to solve the scattering problem when we evaluate the 
vacuum Green's functions, in thermal field theory, one evaluates the thermal Green's functions
(see \cite{1999hepc.conf..187Q} for a review and references).
Just like zero-temperature field theory, the connected 1 PI thermal Green's functions of the field can be found 
from a corresponding generating functional which is the thermal effective action of the field.
The dynamics of the field shall then be governed by the corresponding thermal effective potential.
Given the action of a theory, one can find the Feynman rules to evaluate thermal correlators and hence the thermal 
effective potential.
E.g. for a self interacting scalar field theory, 
the one loop quantum effective potential in thermal state is the sum of two contributions:
$V_{\rm eff}^{\beta}(\phi) = V_{0}(\phi) + V_{1}^{\beta}(\phi)$, where $V_{0}(\phi)$ is the tree level potential.
It turns out that 
\begin{equation}
 V_{1}^{\beta}(\phi) = V^{\rm 1-loop}_{\rm eff}(\phi) + \frac{1}{2 \pi^2 \beta^4} J_B [m^2(\phi) \beta^2] \; ,
\end{equation}
where $V^{\rm 1-loop}_{\rm eff}(\phi)$ is the zero temperature effective potential with $m^2(\phi) = d^2 V_0(\phi)/d \phi^2$.
There is an additional temperature dependent piece 
\begin{equation}
 J_B [m^2(\phi) \beta^2] = \int_0^{\infty} dx x^2 \log [1 - e^{-\sqrt{x^2+\beta^2 m^2}}] \; .
\end{equation}
In the limit of high temperature i.e. $m^2 \beta^2 \ll 1$, the above integral admits a convenient 
expansion.
Similar expressions can be obtained for a theory of a scalar field interacting with a spin half field 
or with a spin one field \cite{1999hepc.conf..187Q}.
Using this, one can obtain the one loop effective potential of the SM Higgs field (in a thermal background)
due to contributions from only the $W$ and $Z$ bosons and the top quark to radiative corrections,
the full one loop thermal effective potential, in the high temperature limit is given by 
\cite{1999hepc.conf..187Q}
\begin{equation} \label{eq:thermal_eff_pot}
V(\phi,T)=D(T^2-T_o^2)\phi^2-ET\phi^3+\frac{\lambda(T)}{4}\phi^4 \; ,
\end{equation}
where the coefficients are given by
\begin{equation}
\label{valord}
D=\frac{2m_W^2+m_Z^2+2m_t^2}{8\phi^2} \; ,
\end{equation}
\begin{equation}
\label{valore}
E=\frac{2m_W^3+m_Z^3}{4\pi \phi^3} \; ,
\end{equation}
\begin{equation}
\label{valort02}
T_o^2=\frac{m_h^2-8B\phi^2}{4D} \; ,
\end{equation}
\begin{equation}
\label{valorb}
B=\frac{3}{64\pi^2 \phi^4}\left(2m_W^4+m_Z^4-4m_t^4\right) \; ,
\end{equation}
%
%
\begin{eqnarray}
\label{valorlamb}
\lambda(T)=\lambda-\frac{3}{16\pi^2
\phi^4}\left(2m_W^4\log\frac{m_W^2}{A_B
T^2} \right. \\ \nonumber 
+ m_Z^4\log\frac{m_Z^2}{A_B T^2} 
\left. - 4m_t^4\log\frac{m_t^2}{A_F T^2} \right) \; ,
\end{eqnarray}
where $\log A_B=5.4076-3/2$ and $\log A_F=2.6351-3/2$.
In the above equations, the rolling Higgs field $\phi$ (which is the same as  
the renormalization scale $\mu$), determines the masses 
such as $m_h^2 = \lambda \phi^2 /2$, $m_W = g_2 \phi/2$, $m_t = y_t \phi/\sqrt{2}$.
Thus, for any renormalization scale $\mu$, we can find the values of $(g_1, g_2, g_3, y_t, \lambda, m^2)$ 
and from them, we can find all the quantities in the above set of equations.

\begin{figure}
  \begin{center}
    \includegraphics[width=.45\textwidth]{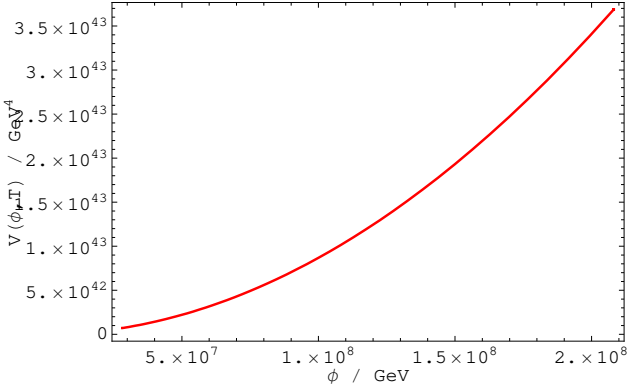}
  \end{center}
\caption{The one-loop thermal effective potential of the SM Higgs field in the field range 
around the instability scale and a temperature of $T = 10^{14}$ GeV. 
The range of field values of this figure and Fig (\ref{potential_zero}) are identical 
but the range of values of the potentials differ by fourteen orders of magnitude. 
Most importantly, the thermal effective potential is positive in the field range corresponding 
to the instability energy scale while the zero-temperature effective potential turns negative. 
This illustrates the fact that around $\phi \sim \mu_{\rm inst}$, the Higgs stability can be 
restored by the temperature caused by Hawking radiation.}
\label{potential_inf_thermal}
\end{figure}

\begin{figure}
  \begin{center}
    \includegraphics[width=.45\textwidth]{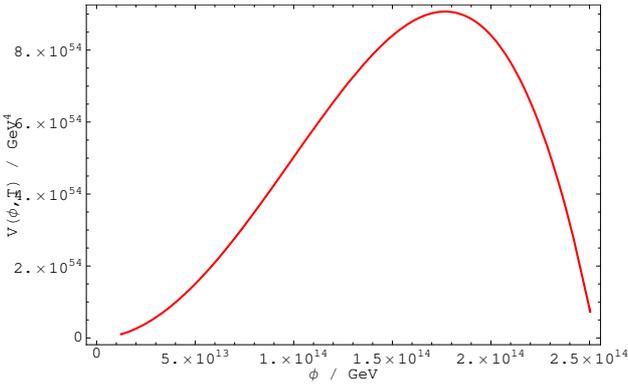}
  \end{center}
\caption{The expected one-loop thermal effective potential of the SM Higgs field in the 
field range around the inflationary Hubble scale and a temperature of $T = 10^{14}$ GeV 
 assuming the validity of high temperature expansion. 
 New physics is certainly required below $H_{\rm inf}$ in order to stabilize this potential
but what this illustrates is that the de-Sitter radiation essentially pushes the instability scale 
from $\mu_{\rm inst} = 10^{8}$ GeV to $H_{\rm inf} = 10^{14}$ GeV.
}
\label{potential_inf_thermal_large}
\end{figure}

Before we proceed, it is worth understanding when the above expressions are valid. 
The high temperature expansion is valid whenever $m^2 (\phi_c) \beta^2 \ll 1$. 
If we choose a value of $\beta$ and if we are interested in a chosen range of $\phi_c$ 
values, then we can find the corresponding $m^2 (\phi_c) \beta^2$ and check whether 
$|m^2 (\phi_c)| \beta^2 \ll 1$ or not. 
Using the tree level Higgs potential, we find that
\begin{equation}
 \beta^2 m^2(\phi) = \beta^2 m^2(\bar{\mu}) + 12 \lambda(\bar{\mu}) \beta^2 \phi^2 \; ,
\end{equation}
where the $m^2$ on the RHS is the mass term which turns up the tree level Higgs potential. 
Since $T \sim H_{\rm inf}$, and we are interested in the range of field values around 
$\phi \sim \mu_{\rm inst}$, it is clear that the high temperature expansion is valid in 
the situation of our interest.
The corresponding one-loop thermal effective potential found from Eq (\ref{eq:thermal_eff_pot}) 
and is shown in Fig (\ref{potential_inf_thermal}). 
As Eq (\ref{eq:thermal_eff_pot}) suggests, for the field values around $\mu_{\rm inst}$, 
the thermal effective potential is governed by $T^2 \phi^2$, so that the stability is restored 
because of large thermal corrections to the mass of Higgs.
On the other hand, for field range of the order of $10^{14}$ GeV (when the quartic term dominates 
over quadratic term), the thermal effective potential begins to turn negative (see Fig (\ref{potential_inf_thermal_large})) 
and we certainly need new physics around this scale to keep the thermal potential positive. 
But around such field values, the high temperature expansion is no longer valid, although correcting for 
this does not change the order of magnitude of numbers in Fig (\ref{potential_inf_thermal_large}). 
Recall that new physics is anyway expected around this scale in order to have successful inflation. 
What we have thus shown is that (i) the Higgs instability problem during 
inflation in fact implies that no new physics is needed till an energy scale equal to the Hubble 
parameter during inflation, and (ii) because of the effect of de-Sitter radiation, the instability scale 
essentially shifts from $\mu_{\rm inst}$ to $H_{\rm inf}$.

{\it Improvement in survival probability}:
Let us suppose that the scalar potential of the BSM scenario valid below Planck scale is 
\begin{equation}
 V(\phi_i,\phi_h) = V_1(\phi_i) + V_2(\phi_h) + V_3(\phi_i,\phi_h) \; ,
\end{equation}
where $\phi_i$ is the inflaton and $\phi_h$ is the SM Higgs. We assume that $V_3$ is very small 
as compared to $V_1$ and $V_2$ because otherwise the SM Higgs potential will be modified at large
$\phi_h$, potentially solving the Higgs instability problem. 
During inflation, $V_1 (\phi_i) \gg V_2 (\phi_h)$ so that the SM Higgs field can be treated as a free field 
(as we did previously).
The probability distribution function of $\phi_h$ is then a Gaussian with mean $ = 0$
and variance $= (H_{\rm inf}/2\pi)^2$. Without taking into account the de-Sitter radiation, 
the probability that after inflation, the SM Higgs is found with $\phi_h < \mu_{\rm inst} = 10^{-6} H_{\rm inf}$ 
is then given by 
\begin{equation}
2 \int_{0}^{10^{-6}} \frac{dx}{\sqrt{2 \pi}}  e^{-x^2/2} = 7.97 \times 10^{-7}  \; ,
\end{equation}
which is too low. 
On the other hand, if one takes into account the effects of de-Sitter radiation, since the instability scale gets 
pushed to $H_{\rm inf}$, the probability of survival is found by integrating the Gaussian from 
$- H_{\rm inf}$ to $+ H_{\rm inf}$ i.e.
\begin{equation}
2 \int_{0}^{1} \frac{dx}{\sqrt{2 \pi}}  e^{-x^2/2} = 0.68  \; .
\end{equation}
Even when we correct Fig (\ref{potential_inf_thermal_large}) for the non-validity of high 
temperature expansion, this number still stays of order 0.5 (the exact number can be worked 
out by redoing all the above without assuming the validity of high temperature expansion).
Thus, taking into account the effects of de-Sitter radiation improves the probability that a given Hubble volume 
will survive the instability in Higgs potential.


{\it Summary}:
It is well known that at a high temperature, broken symmetries get restored. 
In the problem we studied, the Gibbons-Hawking temperature due to de-Sitter radiation 
(as seen by any geodesic observer in quasi de Sitter space)
ensures that the stability of effective potential of SM Higgs gets restored in the 
field range around $\phi \sim \mu_{\rm inst}$.
This is seen clearly if we compare Fig (\ref{potential_zero}) with Fig (\ref{potential_inf_thermal}).
Note that the thermal effective potential shown in Fig (\ref{potential_inf_thermal}) is
positive and has no instability, moreover, if we assume the Hubble parameter during inflation to be 
$H_{\rm inf} \sim 10^{14}$ GeV, 
then, the thermal effective potential of the SM Higgs field in 
the field range around $\phi \approx\mu_{\rm inst}$ (the field value at which $\lambda$ turns negative) 
is ${\cal O}(T^2 \phi^2) \sim 10^{43} {\rm GeV}^4$, which is fourteen orders of magnitude bigger than 
$\lambda(\phi) \phi^4 \sim 10^{29} {\rm GeV}^4$, the effective potential of the SM Higgs in the same 
field range. Moreover, $V_{\rm eff} (\phi,T) \approx 10^{43} {\rm GeV}^4$ is too small compared to 
$V_{\rm inf} \approx 10^{64} {\rm GeV}^4$, the inflaton potential energy density, so that these thermal 
effects in Higgs do not affect the inflationary dynamics. It is thus clear that even if the energy scale of 
inflation is a few orders of magnitude lower than $10^{16}$ GeV, stability is still maintained 
due to Gibbons-Hawking temperature of inflationary quasi de-Sitter spacetime.

If we look at the thermal effective potential of the SM Higgs around $H_{\rm inf}$, i.e.
Fig (\ref{potential_inf_thermal_large}), we find that the de-Sitter radiation has essentially caused the 
instability scale to shift from $\mu_{\rm inst}$ to $T$ which is of order $H_{\rm inf}$.
Because of this, the probability that the SM Higgs is found in the stable part of the 
potential during inflation is quite significant. 
We have thus shown how the considerations of de-Sitter radiation help alleviate the Higgs instability 
problem during inflation.
At the end of inflation the universe becomes radiation dominated by the decay of the inflaton. 
The radiation era has no permanent horizon and there is no Gibbons-Hawking temperature 
associated with the radiation era. The Higgs potential in the radiation era can have interesting 
consequences such as leptogenesis as has been explored in \cite{Kusenko:2014lra}. 

\noindent {\bf Acknowledgment:} 
 G.G. would like to thank Namit Mahajan, Subrata Khan and Raghavan Rangarajan for discussions
 at various stages of the work.

\bibliographystyle{elsarticle-num}
\bibliography{Higgs,BICEP,biblio_inf_tmp}

\end{document}